\documentclass[fleqn,usenatbib]{mnras}

\usepackage{newtxtext,newtxmath}

\usepackage[T1]{fontenc}
\usepackage{ae,aecompl}

\usepackage{epsfig, color}
\usepackage{graphicx}	
\usepackage{amsmath}	
\usepackage{amssymb}	
\usepackage[compact]{titlesec}

\def\ltsima{$\;\buildrel < \over \sim \;$} 
\def\simlt{\lower.5ex\hbox{\ltsima}}
\def\gtsima{$\; \buildrel >\over\sim\;$}
\def\simgt{\lower.5ex\hbox{\gtsima}}

\title[Jet lobes in realistic clusters]{AGN jet feedback on a moving mesh:
  lobe energetics and X-ray properties in a realistic cluster environment} \author[]{Martin A.
  Bourne$^{1,2,\star}$, Debora Sijacki$^{1,2}$ and Ewald Puchwein$^{1,2, 3}$\\ 
  $^{1}$ Institute of Astronomy, University of Cambridge, Madingley Road, Cambridge, CB3 0HA, UK\\ 
  $^{2}$ Kavli Institute for Cosmology, University of Cambridge, Madingley Road, Cambridge, CB3 0HA, UK\\ 
  $^{3}$ Leibniz-Institut f{\"u}r Astrophysik Potsdam (AIP), An der Sternwarte 16, D-14482 Potsdam, Germany\\
  $^{\star}$ {E-mail:~} {\rm mabourne@ast.cam.ac.uk} }

\date{}

\pubyear{2019}

\begin{document}
\label{firstpage}
\pagerange{\pageref{firstpage}--\pageref{lastpage}}
\maketitle

\begin{abstract} 
Jet feedback from active galactic nuclei (AGN) harboured by brightest cluster galaxies is expected to play a fundamental role in regulating cooling in the intracluster medium (ICM). While observations and theory suggest energy within jet lobes balances ICM radiative losses, the {\it modus operandi} of energy communication with the ICM remains unclear. We present simulations of very high-resolution AGN-driven jets launching in a live, cosmological galaxy cluster, within the moving mesh-code {\sc arepo}. As the jet propagates through the ICM the majority of its energy, which is initially in the kinetic form, thermalises quickly through internal shocks and inflates lobes of very hot gas. The jets effectively heat the cluster core, with $PdV$ work and weather-aided mixing being the main channels of energy transfer from the lobes to the ICM, while strong shocks and turbulence are sub-dominant. We additionally present detailed mock X-ray maps at different stages of evolution, revealing clear cavities surrounded by X-ray bright rims, with lobes being detectable for up to $\sim 10^8$~yrs even when magnetic draping is ineffective. We find bulk motions in the cluster can significantly affect lobe propagation, offsetting them from the jet direction and imparting bulk velocities that can dominate over the buoyantly-rising motion.  

\end{abstract}




\section{Introduction}

Feedback, in the form of relativistic jets launched by an accreting supermassive black hole (SMBH), is thought to be critical in regulating the heating and cooling of the intracluster medium \citep[ICM, see e.g.][for reviews]{McNamara2007, Fabian12}. The X-ray cavities produced as a result of lobe inflation \citep[see e.g.][for well known examples]{FormanEtAl07, FabianEtAl2011} seem to be ubiquitous within cool core clusters \citep{DunnFabian08,Fabian12,Hlavacek-LarrondoEtAl12} and exhibit a clear correlation between the estimated lobe energy and the ICM radiative cooling losses. However, while the energetics marry up well, there is still ongoing debate over how exactly the jet energy is effectively and largely isotropically communicated to the ICM. Given that a number of mechanisms and physical processes, such as shocks, sound waves, turbulence, mixing, thermal conduction and cosmic rays \citep[see e.g.,][]{ChurazovEtAl02, McNamara2007, ZhuravlevaEtAl2014, Soker16, YangReynolds16a, EhlertEtal18}, could be important, this issue remains unresolved. Yet this is of fundamental importance for understanding galaxy formation as AGN-driven jet feedback is one of the key physical processes invoked to explain the properties of all massive galaxies. 

Numerical simulations of jets provide an excellent test bed to address this problem given its highly non-linear and complex nature. However, many previous works typically focus their efforts on modelling either the cosmological cluster evolution with simplified AGN heating models \citep[for recent works see e.g.][]{DuboisEtAl10, McCarthyEtAl17, BarnesEtAl18, HendenEtAl18, TremmelEtAl19} or the detailed AGN jet injection in isolated setups that lack realistic thermodynamical properties \citep[e.g.][]{HardcastleKrause13, YangReynolds16b, BourneSijacki17, WeinbergerEtAl17}. To date only a few, restricted studies that follow in detail the jet-inflation of cavities in a full cosmological environment exist in the literature \citep[e.g.][]{HeinzEtAl06, MorsonyEtal2010, MendygralEtAl12}. Therefore, we present high resolution simulations of a live, cosmological galaxy cluster using our recently developed jet feedback scheme \citep{BourneSijacki17}, within the moving mesh-code {\sc arepo} \citep{SpringelArepo2010}. Unlike previous works our simulations also include models for radiative cooling and heating, star formation, supernovae feedback as well as SMBH accretion and feedback based on the Illustris simulation suite \citep{NelsonEtAl15}. Additionally, we employ here specialised refinement criteria to ensure that the AGN-driven lobes are modelled at very high resolution at all times. This allows us to follow their initial inflation and subsequent evolution as well as development of fluid instabilities, turbulence and surrounding gas mixing and entrainment with unprecedented accuracy within a fully self-consistent cosmological cluster simulation.

\section{Numerical method}

The simulations presented here were performed using the moving mesh-code {\sc arepo}, and a more detailed account of the models used will be presented in a follow up paper (Bourne et al., in prep.). In brief, adopting the {\it Wilkinson Microwave Anisotropy Probe} 9-year cosmology \citep{HinshawEtAl13} with Hubble parameter $h = 0.704$, we evolved a cosmological zoom-in simulation of a $M_{\rm 200,c} = 4.14\times 10^{14}$~h$^{-1}$~M$_{\odot}$\footnote{$M_{\rm 200,c}$ is the total mass within a sphere of radius $R_{200,c}$, defined as the radius in which the mean density is equal to 200 times the critical density of the Universe.} galaxy cluster to a redshift of $z\simeq 0.1$ using sub-grid models for gas radiative processes, interstellar medium (ISM) and SMBH physics almost identical to those employed in the original Illustris project, bar a change to the radio-mode AGN feedback model used. Specifically, it was found that the chosen parameters for Illustris resulted in feedback that ejected too much gas from galaxy groups and clusters \citep{GenelEtAl14}. In this work we have therefore adopted a more gentle but more frequent feedback in the radio-mode\footnote{In the model of \cite{SijackiEtAl07}, radio-mode feedback operates by placing hot bubbles of gas (to mimic radio lobes) around accreting BHs. This is done whenever a BH's mass increases by a fraction $\delta_{\rm BH}$, with the bubble energy determined by the rest mass energy associated with this growth. In this work we have assumed $\delta_{\rm BH}=0.015$, which is an order of magnitude smaller than in Illustris and results in bubbles being injected more often but with a lower energy content.}. With this model we find that the total gas mass within $R_{500,c}$ accounts for $14\%$ of $M_{\rm 500,c}$ at $z=0.1$ for our cluster. 

We took the resulting cluster at this redshift as our initial conditions; ``traditional'' SMBH feedback models \citep[for further details see][]{SijackiEtAl2015} were switched off and we instead employed the kinetic jet feedback model presented in \cite{BourneSijacki17}. For this work we assume a fixed gas inflow rate near the black hole of $\dot{M}_{\rm in}=2\times 10^{-4}\dot{M}_{\rm Edd}$, where $\dot{M}_{\rm Edd}$ is the Eddington rate, with half of the inflowing gas entrained in the jet. This sets the jet mass loading factor $\eta_{\rm jet}=\dot{M}_{\rm jet}/(\dot{M}_{\rm in}-\dot{M}_{\rm jet})=1$, which determines the jet mass injection rate 
\begin{equation}
\dot{M}_{\rm jet}=\dot{M}_{\rm in}{\eta_{\rm jet}}/(1+\eta_{\rm jet})
\label{eq:mdot_jet}
\end{equation}
and power
\begin{equation}
    P_{\rm jet}=\epsilon_{\rm jet}\epsilon_{\rm r}\dot{M}_{\rm in}c^{2}/(1+\eta_{\rm jet}),
\label{eq:Edot_jet}
\end{equation}
once the radiative efficiency ($\epsilon_{\rm r}=0.2$) and jet coupling efficiency ($\epsilon_{\rm jet}=1$) are assumed. The jet is active for $20$~Myrs with a power of $\simeq 3.9\times 10^{44}$~erg~s$^{-1}$, given a simulated BH mass of $M_{\rm bh}=2.17\times 10^{10}$~h$^{-1}$~M$_{\rm \odot}$.

\begin{figure*}
\psfig{file=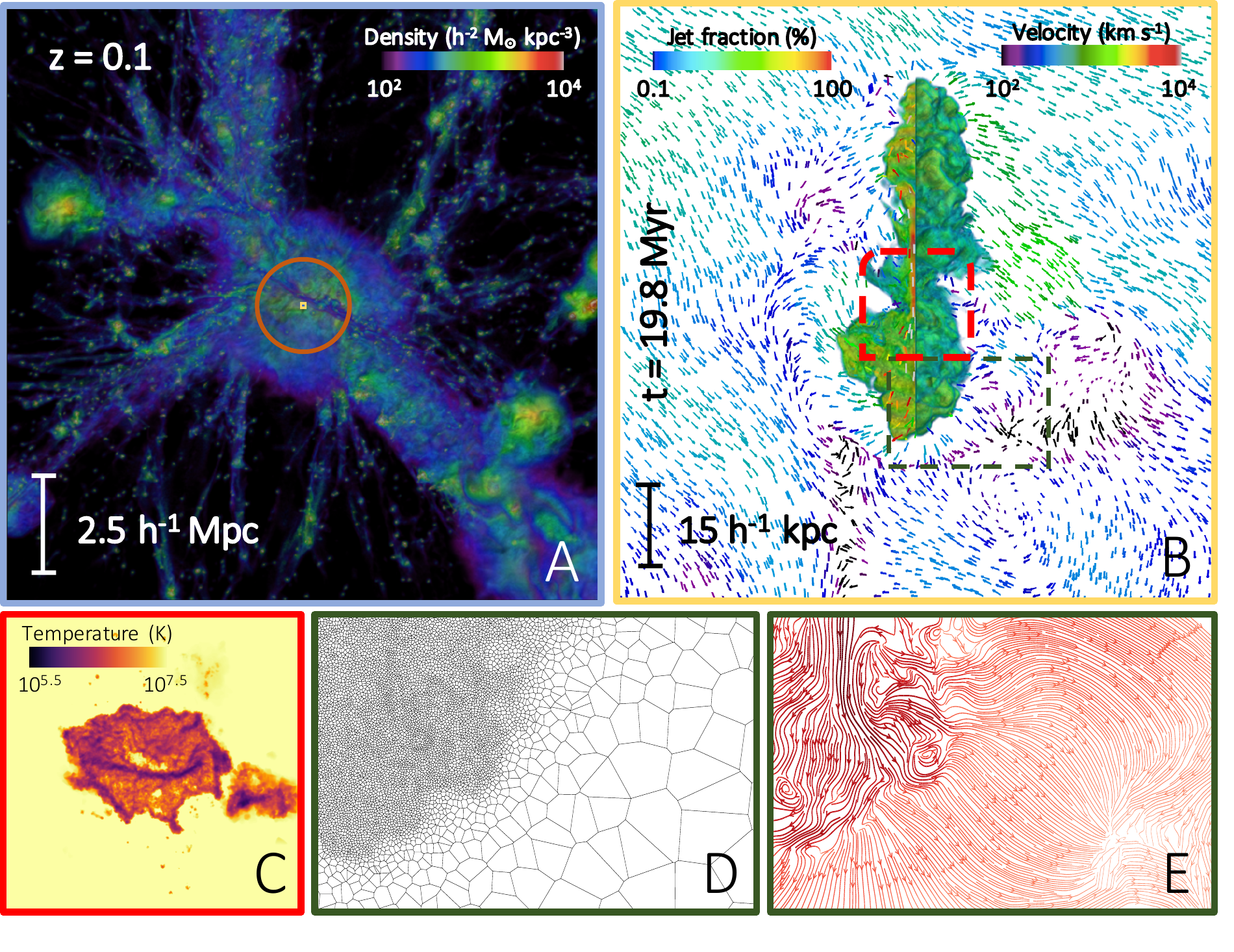, width=0.92\textwidth,angle=0}
\caption{Panel A shows a volume-rendered image of the gas density in a $L=15$~h$^{-1}$~Mpc box centered on the main cluster. Panel B shows the volume-rendered jet material as well as the gas velocity field (arrow vectors) in the central $110$~h$^{-1}$~kpc region. Panel C shows a mass-weighted temperature projection through the central $20$~h$^{-1}$~kpc of the cluster, highlighting a warped cold disc-like structure. Finally, panels D and E show a 2D Voronoi mesh reconstruction and a velocity streamline map of the lower-right lobe-ICM interface, respectively.}
\label{fig:main}
\end{figure*}

The jet is injected into a cylinder whose volume is minimised for the conditions $n_{\rm cell}^{\rm t/b}\geq~10$ and $M_{\rm cyl}\geq~10^{4}$~$h^{-1}~$M$_{\rm \odot}$, where $n_{\rm cell}^{\rm t/b}$ is the number of cells within the top/bottom half of the cylinder and $M_{\rm cyl}$ is the total gas mass within the whole cylinder. Assuming the kinetic energy injection scheme of \cite{BourneSijacki17}\footnote{This scheme does not impose a fixed jet velocity, instead the velocity a cell within the injection cylinder achieves depends on the energy injected into the cell, it's kernel weight and it's mass. The kernel function leads to an outward positive velocity gradient. In essence, this results in cells being ``accelerated'' along the jet cylinder from velocities of $\sim$ a few$\times 10^3$ km s$^{-1}$ until they reach velocities $\sim 0.15$c, at which point they leave the cylinder and are ``launched'' into the jet. Another side effect of this weighting is to curb the occurrence of internal shocks within the injection cylinder itself.}, jet mass, energy and momentum are injected into the cells within each half cylinder , with the momentum directed along the $z$-axis. These quantities are weighted by the kernel function
\begin{equation}
W_{\rm J}(r, z)\propto V_{\rm cell}{\exp\left(-\frac{r^{2}}{2r_{\rm Jet}^{2}}\right)|z|},
\label{eq:kernel}
\end{equation}
where $r_{\rm Jet}$ is the cylinder radius, $V_{\rm cell}$ is the cell volume and $(r, z)$ gives the cell position in cylindrical polar coordinates. Additionally, we include an advective tracer that is set to $f_{\rm jet}=1$ for cells in the cylinder. To achieve sufficiently high resolution close to the central BH, the super-Lagrangian refinement techniques of \cite{CurtisSijacki15} and \cite{BourneSijacki17} are activated as well as additional refinements on the cell volume within jet lobes and accounting for neighbouring cell volumes \citep[similar to][]{WeinbergerEtAl17}. The whole simulated zoom-in region spans $\sim 30$~h$^{-1}$~cMpc across, with a target cell mass of $m_{\rm cell}^{\rm target}=1.37\times 10^{7}$~h$^{-1}$ M$_{\rm \odot}$. However, within the cluster centre cell masses and sizes can be as small as $\sim$~a~few~h$^{-1}$~M$_{\rm \odot}$ and $\sim 10$~h$^{-1}$~pc, respectively, while the typical spatial resolution of cells within the jet lobes is $\sim 100$~h$^{-1}$~pc. The dynamic range spanned between the jet lobe material and the ICM is highlighted in Panel D of Figure~\ref{fig:main}, which shows a 2D reconstruction of the Voronoi mesh.

\section{Results}

\subsection{Overview}
\label{sec:overview}
Panel A of Figure~\ref{fig:main} shows a large-scale view of volume-rendered gas density. The zoom-in cluster located at the centre of the image lies at the intersection of several rich filaments that are permeated by numerous smaller groups and galaxies. It has a virial radius of $R_{200,c}=1178$~h$^{-1}$~kpc (orange circle), which encloses a gas fraction of $M_{\rm g}/M_{200,c}=0.15$, and was chosen as it exhibits no recent AGN activity. The central BH, which acts as the ``anchor point'' of the jet feedback scheme, is surrounded by a $\sim4\times~10^{10}$~h$^{-1}$M$_{\rm \odot}$ disc-like structure of cold gas as shown by the projected temperature map in panel C. The morphology of this central cold gas shows a number of departures from a regular disc structure including a somewhat warped shape. Additionally, the plane of the disc is not perpendicular to the jet direction and the BH sits just above cold material. Once the jet is launched, this results in an interaction between the jet and cold material that opens up the central hole seen in the panel and impacts the resulting lobe morphology (see  Section~\ref{sec:mock}). Similar structures have been observed in the centres of a number of galaxy clusters \citep[e.g.][]{HamerEtAl16}.

When active, the high velocity jet ($v_{z}\simgt~0.1$~c) inflates lobes of hot gas ($T \sim 10^{10}$~K). The lobe structure is shown in panel B, which zooms in to the central region of the cluster $19.8$~Myrs after the jet is switched on. The jet material is illustrated by a volume rendering of $f_{\rm jet}$, where the right hand side shows the surface structure of the lobes, while on the left hand side we have made a cut to show the internal structure at the midplane. The lobe material itself is stirred by the jet, resulting in a small turbulent contribution to the lobe energy budget (see Figure~\ref{fig:lobe_energy_evo} for further details), while the rugged nature of the lobe surface is a result of instabilities driven along the ICM-jet lobe interface. The turbulent nature of the lobe material can be seen more clearly in panel E, which shows the velocity field streamlines. 

Beyond the immediate lobe structure extends the cocoon of swept-up and heated ICM material bounded by a discontinuity in the velocity field that is distorted in places by the ICM `weather'. This can be seen in the ICM velocity field in the plane of the jet lobes as depicted by the coloured arrows in panel B. This weather ultimately acts to displace the top lobe from its original trajectory while a substructure moving towards the cluster centre from the lower-right, will interact strongly with the bottom lobe and aid in mixing the jet material with the ICM. 

\begin{figure*}
\psfig{file=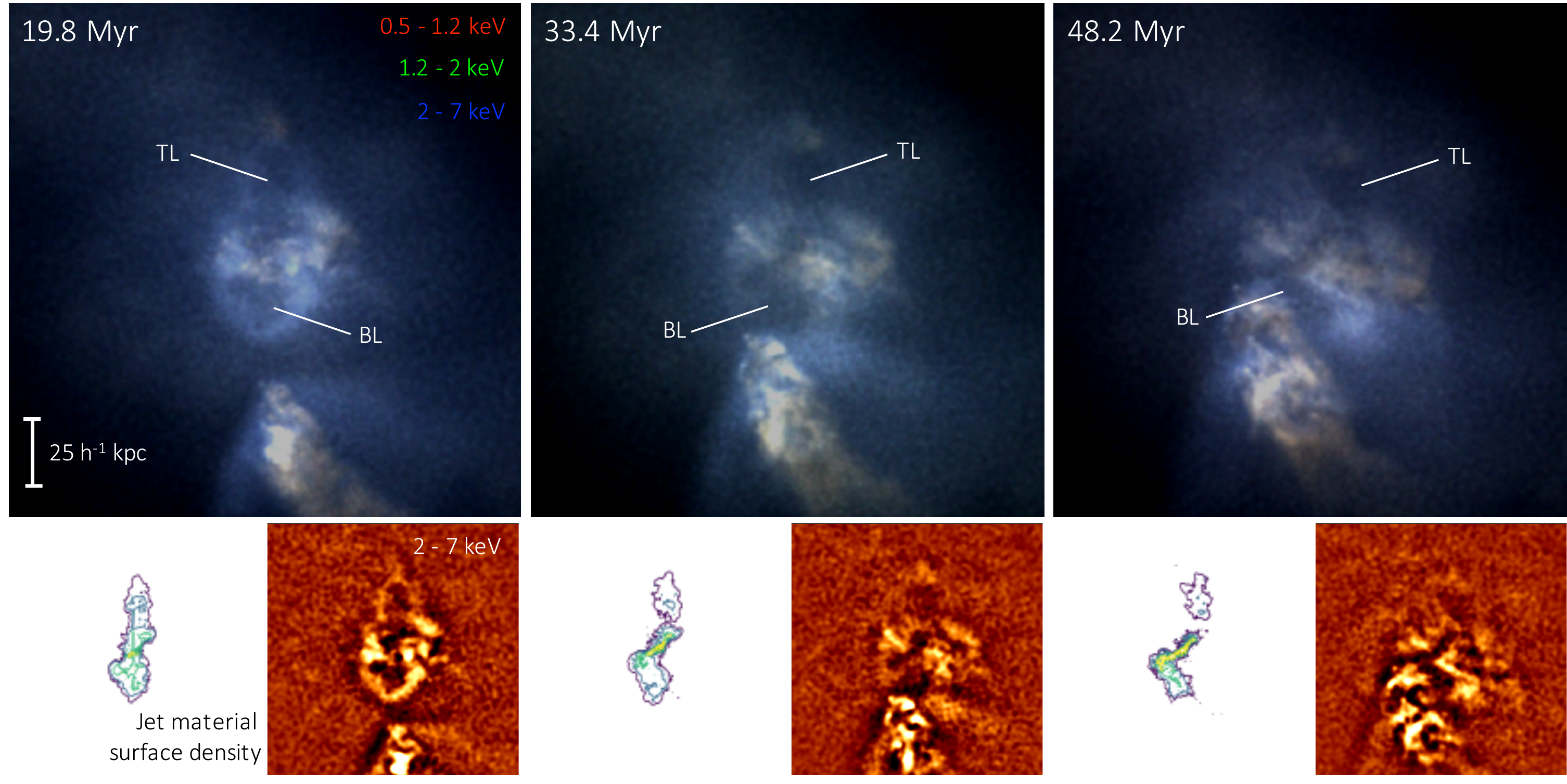, width=0.9\textwidth,angle=0}
\caption{The top row shows RGB composite X-ray images of the cluster center at different stages of lobe evolution, where the energy bands are $0.5-1.2$~keV (red), $1.2-2$~keV (green) and $2-7$~keV (blue), with the top (TL) and bottom (BL) lobes labelled. Below each of the top panels are two additional panels showing contours of the jet material surface density and unsharp-masked images of the $2-7$~keV band on the left and right, respectively.}
\label{fig:mock_xray}
\end{figure*}

\subsection{Mock X-ray images}
\label{sec:mock}

We have produced mock X-ray images of the jet cavities using the {\sc pyXSIM} package \citep{ZuHoneEtAl14}, assuming a fixed metallicity of $0.3$ solar and that the cluster is at the same redshift as the Perseus cluster \citep{FabianEtAl06}. The use of an effective equation of state for star forming gas and the lack of molecular cooling in our simulations means that we are unable to reliably capture the thermal properties of cold dense gas. Therefore, as in \citet{RasiaEtAl12} we apply a temperature-density cut that excludes gas with $T_{\rm keV} < 3\times~10^6\rho^{0.25}_{\rm cgs}$ when generating the X-ray photons. The gas cut by this method would likely exist in a colder phase than modelled in our simulations and hence is unlikely to actually be observed in the X-ray band. The top row of Figure~\ref{fig:mock_xray} shows RGB composite images of the $0.5-1.2$(R), $1.2-2$(G) and $2~-~7$(B)~keV energy bands at $19.8$, $33.4$ and $48.2$~Myr since the jet is switched-on (note jet switches-off at $20$~Myr). The images were smoothed on a scale of $\sim 4$~h$^{-1}$~kpc with a Gaussian filter to reduce noise. The bottom row shows surface density contours of jet material and unsharp-masked images for the $2-7$~keV energy band at the corresponding times. 

The cavities are clearly visible in the images produced at $19.8$~Myr, with X-ray bright rims prominent across all energy bands in the RGB image, particularly for the bottom lobe. They are also picked up in the unsharp-masked image. Such rim features are seen in numerous observations of cool core galaxy clusters such as Perseus \citep[e.g.,][]{FabianEtAl06}. An asymmetry between the top and bottom lobes, somewhat similar to that seen in Abel 4059 \citep{HeinzEtAl02} and Abell 2052 \citep{BlantonEtAl01}, is present at this time and we note this is due to the interaction of the jet with the central cold disc. The bottom jet appears to interact more strongly with the cold gas, which impedes its progress, while the top jet has a clearer path, primarily interacting with hot ICM gas and hence being able to propagate further (see also the discussion in Section \ref{sec:overview}). This also explains, why the bottom lobe rim appears brighter in the X-rays. In fact, while the top lobe rim is prominent in the $2-7$ keV band, we find that the bottom lobe is clearly visible in the lower energy bands too, as it contains cooler material. 

Once the jet has switched off the lobe structure becomes less obvious, without prominent rims, although it can still be detected in the RGB images as depressions in the X-ray emission. The top lobe appears to flatten as it ages, similar to observed relic lobes \citep[e.g.,][]{ReynoldsEtAl05} and the cluster weather dominates over buoyancy; pushing the top lobe to the right, giving the impression that the jet direction was not aligned with the z-axis. Additionally, the motion of a cold substructure coming from the lower-right can be seen predominantly in the soft band. Similar features have been seen in X-ray observations of e.g. Abell 2142 \citep{EckertEtAl17} and ESO 317-001 in Abell 3627 \citep{SunEtAl10} as galaxies fall into clusters, albeit at larger radii than in our simulated case. This structure interacts strongly with the bottom lobe, compressing it (as seen in the right hand panels), before completely disrupting it. In the space of almost $30$ Myr, the dynamic nature of the cluster leads to very different looking environments, from the archetypal cavity structure seen in many cool core clusters to a much messier structure, akin to the cluster $2a 0335$ \citep{SandersEtAl09}, in which it becomes more difficult to definitively identify the location of cavities visually even though the lobes still retain $40 \%$ of the cumulative jet energy.

\subsection{Lobe energetics}

\begin{figure*}
\psfig{file=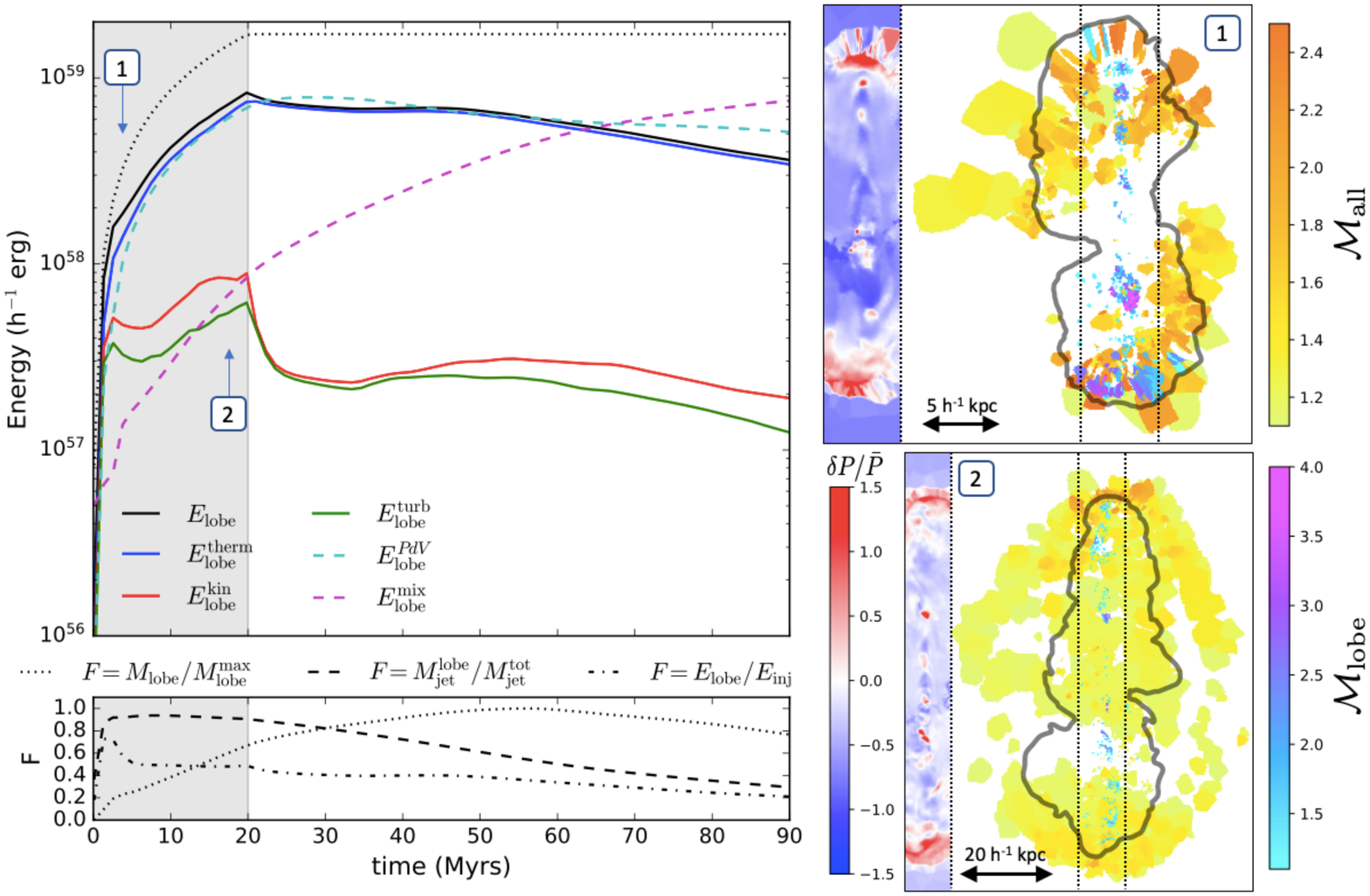,width=0.98\textwidth,angle=0}
\caption{{\it Left hand side:} Evolution of jet lobe energy content (solid black line) is shown in the {\it top panel}, decomposed into the thermal, kinetic and turbulent component. The grey shaded region indicates the period over which the jet is active. The total injected energy is shown by the dotted black line. Dashed cyan and magenta lines show estimated $PdV$ work and mixing losses, respectively. The {\it bottom panel} shows the evolution of the total lobe mass normalized to its maximum value (dotted),  injected jet mass within the lobe normalized to the total injected mass (dashed) and lobe energy normalized to the total injected energy (dot-dashed). {\it Right hand side:} Dissipation-weighted projections of shock Mach numbers are shown by the yellow/orange maps at $t=3.79$~Myr and $17.36$~Myr in panels 1 and 2, respectively. Additionally, shocks occurring within jet lobe material are overlaid in blue/pink, with pressure fluctuations produced by internal shocks along the jet also shown by blue/red maps.}
\label{fig:lobe_energy_evo}
\end{figure*}

In the following, we define lobe material as cells with $f_{\rm jet}>f_{\rm jet}^{\rm thresh}=10^{-2.5}$ and additionally exclude star forming cells, which are those with $n>0.26$ h$^2$ cm$^{-3}$. The value of $f_{\rm jet}^{\rm thresh}$ is similar to previous works \citep[e.g.][]{HardcastleKrause13, WeinbergerEtAl17, BourneSijacki17,YangReynolds16a} that employ typical values of $0.01-0.001$. In our work this represents the point at which the gas density/temperature start to decrease/increase respectively, transitioning from ICM like values to lobe like values. This can be understood very simply if we decompose a cell's internal energy into the contributions from the ICM and lobe material such that $U_{\rm therm}^{\rm cell}=m_{\rm cell}\times[f_{\rm jet}u_{\rm jet} + (1-f_{\rm jet})u_{\rm ICM}]$. Assuming $f_{\rm jet}\ll 1$ we find that the cell energy transitions from being lobe dominated to ICM dominated when $f_{\rm jet}\leqslant u_{\rm ICM}/u_{\rm jet}\sim T_{\rm ICM}/T_{\rm jet}$, which for typical temperatures in our simulations corresponds approximately to $f_{\rm jet}^{\rm thresh}$.

The evolution of the jet lobe energy content is presented in the left hand panels of Figure~\ref{fig:lobe_energy_evo} showing the total, thermal, kinetic and turbulent components of the lobe energy. Additionally, the cumulative jet-injected energy is shown by the black dotted line. Similar to other works \citep[e.g.][]{WeinbergerEtAl17, EhlertEtal18}, the turbulent energy is estimated by defining the turbulent velocity of each cell by subtracting the mean velocity vector of the relevant lobe from the cells velocity vector. Further, to avoid contamination from the high bulk velocity of the jet itself, cells with $|v_{z}|>0.1c$ are neglected when estimating the turbulent velocity. We note that while more sophisticated methods of estimating the turbulent component of the velocity field could have been employed; such as velocity field decomposition \citep[e.g.][]{RyuEtAl08, ZhuEtAl10, ReynoldsEtAl2015}, fixed scale filtering \citep[e.g.][]{DolagEtAl2005, VazzaEtAl09,Valdarnini2011} or multi-scale filtering \citep[e.g.][]{VazzaEtAl12, VazzaEtAl17, BourneSijacki17, Valdarnini19}, given the small fraction of injected energy retained in the kinetic form within the lobes (see below), the turbulent component can only be comparable or less than this i.e. also a small contribution to the total energy budget.

The difference between the total lobe energy and cumulative injected energy represents the energy transferred to the ICM via various physical processes. We estimate cumulative lobe losses due to $PdV$ work and mixing by integrating over $\Delta E_{\rm lobe}^{PdV}=\overline{P}_{\rm lobe}\times\Delta V_{\rm lobe}$ and $\Delta E_{\rm lobe}^{\rm mix} = \overline{\epsilon}_{\rm lobe}\times \Delta M_{\rm jet}^{\rm mix}$, respectively, where $\overline{P}_{\rm lobe}$ and $\overline{\epsilon}_{\rm lobe}$ are averages of the lobe pressure and lobe energy per unit mass of jet material\footnote{By jet material we mean the mass injected into the jet through Equation \ref{eq:mdot_jet}, this is different to the jet lobe material defined by $f_{\rm jet}>10^{-2.5}$.}, and $\Delta V_{\rm lobe}$ and $\Delta M_{\rm jet}^{\rm mix}$ are changes in lobe volume and mass of jet material that mixes into the ICM, calculated between consecutive snapshots, respectively. It is also worth mentioning that calculated quantities can depend on the exact choice of $f_{\rm jet}^{\rm thresh}$, i.e. while the total lobe energy content and $PdV$ work estimate increase for smaller values of $f_{\rm jet}^{\rm thresh}$, the estimated mixing decreases. However, based on our discussion above we believe our choice for $f_{\rm jet}^{\rm thresh}$ is well motivated and provided it is small enough our qualitative conclusions are insensitive to its exact value\footnote{We point out that given how well our estimates of the $PdV$ work and mixing account for lobe losses and are thus able to recover the total energy budget, we expect our approximate method is justified.}.

Even though the jet energy is injected almost exclusively in the kinetic form, the majority of this energy rapidly thermalises through shocks, which leads to the thermal energy component dominating the total lobe energy throughout the evolution. The importance of shocks for the process of thermalisation of the jet kinetic energy and inflation of the hot lobes has been highlighted in a number of previous studies \citep[e.g.][]{BourneSijacki17, WeinbergerEtAl17, YangReynolds16a, MartizziEtAl18}. To explore this further here, we use the algorithm of \citet{SchaalSpringel15} to detect shocks produced by the action of the jet and lobe inflation at two distinct times as labelled on the left hand panel of Figure~\ref{fig:lobe_energy_evo}. Dissipation-weighted projections of shock Mach numbers are shown by the yellow/orange maps\footnote{To guard against misclassifying contact discontinuities as shocks, \citet{SchaalSpringel15} require $\Delta\log T\geqslant\log \frac{T_2}{T_1}|_{\mathcal{M}=\mathcal{M}_{\rm min}}$ and $\Delta\log P\geqslant\log \frac{P_2}{P_1}|_{\mathcal{M}=\mathcal{M}_{\rm min}}$, with $\mathcal{M}_{\rm min}=1.3$. While we apply this condition for the majority of our analysis, in order to illustrate the location of weak shocks, we use $\mathcal{M}_{\rm min}=1.1$ for the production of the Mach number projections in Figure~\ref{fig:lobe_energy_evo}.} in the right hand panels of Figure~\ref{fig:lobe_energy_evo}, with the jet lobe footprints indicated by the grey contours. We find that as well as a bow shock at the ends of the cocoon being driven in to the ICM (see discussion below), multiple regions of internal shocks occur along the jet axis. We distinguish shocks occurring within the jet lobes themselves by the overlaid blue/pink maps. We detect internal shocks spanning a range of Mach numbers, and while the upper end of this range typically reaches $\mathcal{M}\sim 3-4$, it can be as high as $\mathcal{M}\sim 8$ in some instances. These internal shocks result in pressure fluctuations along the jet axis, which are shown by the blue/red maps. We expect that while variations in the injected jet velocity can contribute to internal shocks, other physical processes such as the jet interacting with backflows \citep[e.g.][]{BourneSijacki17,CieloEtAl2014, ADS2010} and turbulent motions that occur within the lobes \citep[see also][]{WalgEtAl14} are also important.

Of the residual lobe kinetic energy, which accounts for only $\sim 5\%$ of the injected jet energy, much of it is in turbulence. During lobe inflation, its total energy content (dot-dashed line, lower panel) accounts for $\sim 50\%$ of the cumulative jet energy. Given that radiative cooling is negligible in the lobes, half of the jet energy must be transferred to the ICM during the first $20$~Myr. This is predominantly through $PdV$ work done on the ICM via lobe expansion, which accounts for $\sim 40\%$ of the cumulative jet energy. Interestingly, the lobe enthalpy, $H=E^{\rm lobe}_{\rm therm}+PV$, calculated using the instantaneous lobe $PV$ at $20$~Myr would underestimate the total injected energy by a factor of $\sim 1.4$. 

Similarly to our previous work \citep{BourneSijacki17},  the lobe inflation is initially rapid and drives strong shocks into the ICM both perpendicular to and along the jet direction (see top right hand panel of Figure~\ref{fig:lobe_energy_evo}). At later times only the driving of the bow shock (up to $\mathcal{M}\sim 2-3$) produced in the jet direction persists, while the perpendicular lobe expansion becomes largely sub-sonic resulting in the shock broadening and detachment from the lobes. This is clearly seen in the bottom right hand panel of Figure~\ref{fig:lobe_energy_evo}, where the oval shape outlined by the shocks corresponds to the cocoon boundary. Using the shock finding algorithm of \citet{SchaalSpringel15}, we find that the kinetic energy dissipated via strong shocks ($\mathcal{M} > 1.5$) driven in to the ICM accounts for only a small fraction ($\sim 10\%$) of the $PdV$ work. Therefore, we suggest that much of the $PdV$ work done on the ICM during lobe inflation must go into displacing gas, compressional heating, weak shocks and sound waves. Note that during the lobe inflation phase, mixing is sub-dominant: $\sim 90\%$ of jet material still resides within the lobes by $20$~Myr (dashed line, lower panel) and we estimate that roughly $\sim 5\%$ of the injected energy is transferred to the ICM through mixing by this time. 

However, the picture changes once the jet ceases, with cluster weather becoming important. There is a sharp drop in the kinetic energy once the jet action halts, a slow decline in the thermal energy content of the lobe as it is no longer replenished through shocks, and mixing becomes increasingly more important. $PdV$ losses peak $\sim 7$~Myr after the jet stops, after which they slowly decline, in part due to the bottom lobe being compressed by the incoming substructure, which can be seen in all three RGB images in Figure~\ref{fig:mock_xray} as the bright structure moving up from the lower right. The incoming substructure drives a $\mathcal{M}\sim 2$ bow shock into the ICM, which can be clearly seen in the hard X-ray band (blue). The bow shock compresses and mixes with the bottom lobe, resulting in a small increase in both the thermal and kinetic energies of the lobes at $\sim 40$~Myrs and contributes to the total lobe mass (see dotted line in bottom left hand panel). 

In our previous work \citep{BourneSijacki17} using idealised setups, we found that stirring of the ICM can enhance the rate of mixing and redistribute lobe material. Here the impact is even more pronounced and the potential impact of cluster weather on mixing can be seen by comparing the evolution of the top and bottom lobes. While the cluster weather is able to displace the northern lobe which is pushed to the right (see Figure~\ref{fig:mock_xray}), the interaction of the substructure with the bottom lobe ultimately completely disrupts it. Measuring instantaneous energy loss rates due to mixing for each lobe, we find that the interaction of the substructure with the bottom lobe can result in a factor $\sim 3$ increase compared with the top lobe. Overall, by $\sim 62$~Myr over half of the total jet material has mixed into the ICM and by $90$~Myrs the equivalent of $\sim 44\%$ of the cumulative jet energy has been transferred to the ICM through weather-enhanced mixing. 

\section{Discussion and Conclusions}

We have performed very high resolution simulations of AGN-driven jets in a live cosmological galaxy cluster, finding that the environment and cluster weather can have a significant impact on the lobe inflation and evolution \citep[see also][]{HeinzEtAl06, MorsonyEtal2010}, in particular to aid mixing of jet material with the ICM and hence lead to the effective and largely isotropic energy transport. 

Mock X-ray maps of our simulated cluster exhibit many features seen in a number of observed galaxy clusters across different stages of evolution including cavities surrounded by X-ray bright rims and flattening of the cavities as they propagate and age through the cluster core \citep{FabianEtAl06,HeinzEtAl02,BlantonEtAl01,SandersEtAl09}. Due to the asymmetries in the local gas, which forms a cold, rotationally supported disc, initial propagation of the top and bottom cavities are different, but follow the jet injection axis. However, once the jet is switched off the cavities are pushed and deformed by the ICM motions, although they retain more than $40 \%$ of the jet energy for up to $45$~Myrs. 

Interestingly, jet re-orientation has been used as a possible mechanism to explain the angular offset between different generations of cavities within galaxy clusters \citep[e.g.][]{DunnEtAl06,BabulEtAl13}, with a number of simulation works finding that precession of the jet axis \citep[e.g.][]{VernaleoReynolds06,Falceta-GoncalvesEtAl10,LiBryan14}, or rapid re-orientation of the jet axis by hand \citep[e.g.][]{CieloEtAl18} is able to aid in the isotropic heating of the ICM. In this work we find that the cluster weather alone is able to significantly displace lobes from their initial trajectory and could explain the observed distributions of cavities in some clusters \citep[see also][]{SijackiEtAl08, MorsonyEtal2010, BourneSijacki17}. With upcoming observational missions, such as the X-Ray Imaging and Spectroscopy Mission (XRISM), that will be able to make measurements of the ICM kinematics and hopefully shed light on these different processes, it will also be necessary to simulate lobe inflation in realistic galaxy groups and clusters of different masses across cosmic time in order to provide a theoretical comparison.

Similar to previous results in idealised cluster setups \citep[e.g.][]{HardcastleKrause13, HardcastleKrause14, BourneSijacki17, WeinbergerEtAl17}, we find that during lobe inflation approximately half of the jet energy remains in the lobes, with the rest going into the ICM, predominantly through $PdV$ work. Specifically, while we show that strong shocks are not important for directly heating the ICM, they are important within the jet lobes in order to rapidly thermalise the kinetic jet, with similar conclusions being found in other works performed in idealised setups \citep[e.g.][]{YangReynolds16a, MartizziEtAl18}. At later times cluster weather aids mixing which becomes an equally important channel for transferring energy to the ICM. While some works have suggested that mixing due to small-scale instabilities is important \citep[e.g.][]{HillelSoker16b, HillelSoker17}, we previously found this to be ineffective in a hydrostatic environment \citep{BourneSijacki17} and here instead emphasise the importance of cluster weather in displacing and disrupting hot lobe material \citep[see also][]{DuboisEtAl12, HeinzEtAl06, BourneSijacki17}. If magnetic draping or other processes that suppress mixing are largely ineffective, these two channels of energy transfer are sufficient to heat the cluster core and we find that the central cooling time of the ICM remains $\simgt 8$~Gyr for $\sim 45$~Myrs after the jet injection ceases.

\section*{Acknowledgements}
We would like to thank Helen Russell, Stas Shabala, Eugene Churazov and Nick Henden for useful discussions and comments on this work. MAB and DS acknowledge ERC starting grant 638707 and support from the STFC. EP acknowledges support by the Kavli foundation. This research used: The Cambridge Service for Data Driven Discovery (CSD3), part of which is operated by the University of Cambridge Research Computing on behalf of the STFC DiRAC HPC Facility (www.dirac.ac.uk). The DiRAC component of CSD3 was funded by BEIS capital funding via STFC capital grants ST/P002307/1 and ST/R002452/1 and STFC operations grant ST/R00689X/1. The DiRAC@Durham facility managed by the Institute for Computational Cosmology on behalf of DiRAC. The equipment was funded by BEIS capital funding via STFC capital grants ST/P002293/1 and ST/R002371/1, Durham University and STFC operations grant ST/R000832/1. DiRAC is part of the National e-Infrastructure.





\bibliographystyle{mnras}

\bibliography{ArepoJet}




\bsp	
\label{lastpage}
\end{document}